\documentclass[showpacs,preprintnumbers,prl,titlepage,preprint,showkeys,nofootinbib,bibnotes,10pt,twocolumn]{revtex4}
\usepackage{amsmath}
\usepackage{graphicx}
\usepackage{dcolumn}
\usepackage{bm}

\input{tcilatex}

\begin{document}

\title{Ballistic-Transport-Induced Negative Differential Thermal Resistance}
\author{Wei-Rong Zhong$^{1}$}
\email{wrzhong@hkbu.edu.hk}
\author{Ping Yang$^{1}$}
\author{Bao-Quan Ai$^{1,\text{ }2}$}
\author{Zhi-Gang Shao$^{1}$}
\author{Bambi Hu$^{1,\text{ }3}$}
\affiliation{$^{1}$\textit{Department of Physics, Centre for Nonlinear Studies, and The
Beijing-Hong Kong-Singapore Joint Centre for Nonlinear and Complex Systems
(Hong Kong), Hong Kong Baptist University, Kowloon Tong, Hong Kong.}}
\affiliation{$^{2}$\textit{School of Physics and Telecommunication Engineering, South
China Normal University, 510006 Guangzhou, P. R. China.}}
\affiliation{$^{3}$\textit{Department of Physics, University of Houston, Houston, Texas
77204-5005, USA.}}
\date{\today }

\begin{abstract}
Using nonequilibrium molecular dynamics simulations, we study the
temperature dependence of the negative differential thermal resistance that
appears in two-segment Frenkel-Kontorova lattices. We apply the theoretical
method based on Landauer equation to obtain the relationship between the
heat current and the temperature, which states a fundamental interpret about
the underlying physical mechanism of the negative differential thermal
resistance. The temperature profiles and transport coefficients are
demonstrated to explain the crossover from diffusive to ballistic transport.
The finite size effect is also discussed.
\end{abstract}

\maketitle

Negative differential thermal resistance\ (NDTR) is a property of certain
materials connected to two heat baths in which, over certain temperature
ranges, heat current is a decreasing function of the temperature difference
between the two heat baths, \emph{\emph{i}.e., }$dJ/dT<0$, here $J$ and $T$
are the heat current and the temperature, respectively. Li \emph{et al} had
investigated different kinds of systems to reveal that negative differential
thermal resistance is not a happenstance and indeed take place in low
temperature nonlinear lattices for a certain range of parameters \cite{B.Li1}
\cite{B.Li3} \cite{B.Li4} \cite{Segal2}. It is naturally expected that NDTR
effects may lead to an impressive technological innovation and even the
appearance of the thermal transistor, which maybe change our world
thoroughly in future, just like electronic transistor and other relevant
devices have done in the past half-century \cite{B.Li1} \cite{Bardeen} \cite%
{Segal1}.

Although thermal transistor is confirmed by nonequilibrium molecular
dynamics simulations, the practicable counterpart using fabricated materials
is not produced up to now \cite{B.Li1}. As the main physical mechanism of
thermal transistor, negative differential thermal resistance (NDTR) effect
as well as asymmetric heat conductance still need to be understood deeply 
\cite{CWChang}. The phenomenon of NDTR is firstly understood from the
mismatch between the phonon bands of the two interface particles \cite{B.Li1}%
. In Ref. \cite{YZhang} and \cite{ZGShao}, the authors reported that the
phonon bands are independence of the system size but the NDTR disappears for
the large system. Segal has suggested that NDTR shows up when the molecular
is strongly coupled to the thermal baths in a asymmetric system \cite{Segal2}%
. However, the NDTR in the absence of asymmetry is still not clear \cite{Ai}%
. Therefore, the mismatch of the phonon bands and asymmetry are not a real
physical mechanism of the NDTR. To reveal the real physical mechanism of the
NDTR is still one of new and challenging problems about thermal transport.

Macroscopic Fourier's law, $J=-\sigma \nabla T$, that connects heat current
with thermal transport on the microscopic scale, is an empirical law based
on observation, where $\nabla T$ is the temperature gradient and $\sigma $
is the thermal conductivity. It states that the heat current through a
material is proportional to the negative temperature gradient. On the other
hand, according to microscopic thermodynamics, heat current at low
temperature is proportional to the number of phonons \cite{Phonon}. A phonon
is a quantized mode of vibration occurring in a rigid crystal lattice, such
as the atomic lattice of a solid \cite{SolidPhysics}. The Bose-Einstein
probability distribution for phonons, based on statistical mechanics
concepts for thermal equilibrium, determines the number of phonons. Thus,
the heat current is also determined by the Bose-Einstein probability
distribution. In this paper, applying an analytical method from Landauer
equation as well as nonequilibrium molecular dynamics simulations, we reveal
the dependence of the NDTR on the temperature and the crossover of two
thermal transport processes: the ballistic transport and the diffusive
transport \cite{Roy}. We will also investigate the system size dependence of
the NDTR. The temperature profile and the transport coefficient will be
calculated to characterize the ballistic transport and the diffusive
transport.

The nonlinear lattices that we use in this letter consist of two segments,
left segment (L) and right segment (R). Each segment is a Frenkel-Kontorova
(F-K) lattice. Segment L and R are coupled via a spring of constant $K_{int}$%
. The total Hamiltonian of the model is%
\begin{equation}
H=H_{L}+H_{R}+H_{int},
\end{equation}%
and the Hamiltonian of each segment can be written as%
\begin{equation}
H_{M}=\sum_{i=1}^{N_{M}}\left[ \frac{p_{M,i}^{2}}{2m_{M}}+\frac{K_{M}}{2}%
\left( q_{M,i+1}-q_{M,i}\right) ^{2}+\frac{V_{M}}{(2\pi )^{2}}\cos (\frac{%
2\pi }{a}q_{M,i})\right] ,
\end{equation}%
with $q_{M,i}$ and $p_{M,i}$ denote the displacement from equilibrium
position and the conjugate momentum of the $i^{th}$ particle in segment $M$,
where $M$ stands for $L$ or $R$. The parameters $K$ and $V$ are the harmonic
spring constant and the strength of the external potential of the FK
lattice, respectively. We couple the last particle of segment $L$ and $R$
via a harmonic spring. Thus, $H_{int}=\frac{K_{int}}{2}\left(
q_{L,N}-q_{R,N}\right) ^{2}.$ We set $m=a=1,$ $K_{L}=1.0,$ $K_{R}=0.2,$ $%
V_{L}=5.0,$ $V_{R}=1.0,$ $K_{int}=0.05$.

In our simulations we use fixed boundary condition and the chain is
connected to two heat baths at temperature $T_{L}$ and $T_{R}$. We use the
Nos\'{e}-Hoover heat baths and integrate the equations of motion by using
the fourth-order Runge-Kutta algorithm \cite{Nose} \cite{Art} \cite{Press}.
The local temperature is defined as $T_{i}=\left\langle
p_{i}^{2}\right\rangle $. The local heat flux is defined as $%
j_{i}=K_{M}\langle p_{i}(q_{i}-q_{i-1})\rangle $, and the total heat flux is 
$J=Nj$. The simulations are performed long enough to allow the system to
reach a steady state in which the local heat flux is constant along the
chain. For the sake of comparison, we define a heat current ratio, $%
J_{R}=J/J_{\max }$, in which $J_{\max }$ is the maximum heat current under a
fixed temperature $T_{R}$ of the right heat bath. The transport coefficient
is an important quantity for characterizing the transport mode of a thermal
transport process \cite{JSWang1} \cite{JSWang2} \cite{DeyuLi}. The thermal
conductance evaluated as $\sigma =Nj/\triangle T$ represents an effective
transport coefficient that includes both boundary and bulk resistances \cite%
{Livi}.

Figure 1 displays a typical negative differential thermal resistance \cite%
{B.Li1}. When the temperature difference increases (\emph{\emph{i}.e.,} $%
T_{L}$ ($\leq $ $T_{R}$) decreases), the heat current increases firstly and
then decreases. The former is positive differential thermal resistance
(PDTR) and the latter negative differential thermal resistance. It would be
much interesting to show the thermal conductance dependent behavior of NDTR,
which is also presented in Fig.1. It is clearly that the NDTR corresponds to
the ballistic regime and PDTR the diffusive regime.\FRAME{ftbpFU}{4.7729in}{%
3.4549in}{0pt}{\Qcb{The temperature dependence of heat current ratio (blank
squares) and the corresponding temperature dependence of the thermal
conductance (solid circles). The system parameters are $N_{L}=N_{R}=50$, $%
V_{L}=5.0,$ $V_{R}=1.0,$ $K_{L}=1.0,$ $K_{R}=0.2,$ $K_{int}=0.05,$ and\ $%
T_{R}=0.21$.}}{}{fig1.eps}{\special{language "Scientific Word";type
"GRAPHIC";maintain-aspect-ratio TRUE;display "USEDEF";valid_file "F";width
4.7729in;height 3.4549in;depth 0pt;original-width 4.7193in;original-height
3.4091in;cropleft "0";croptop "1";cropright "1";cropbottom "0";filename
'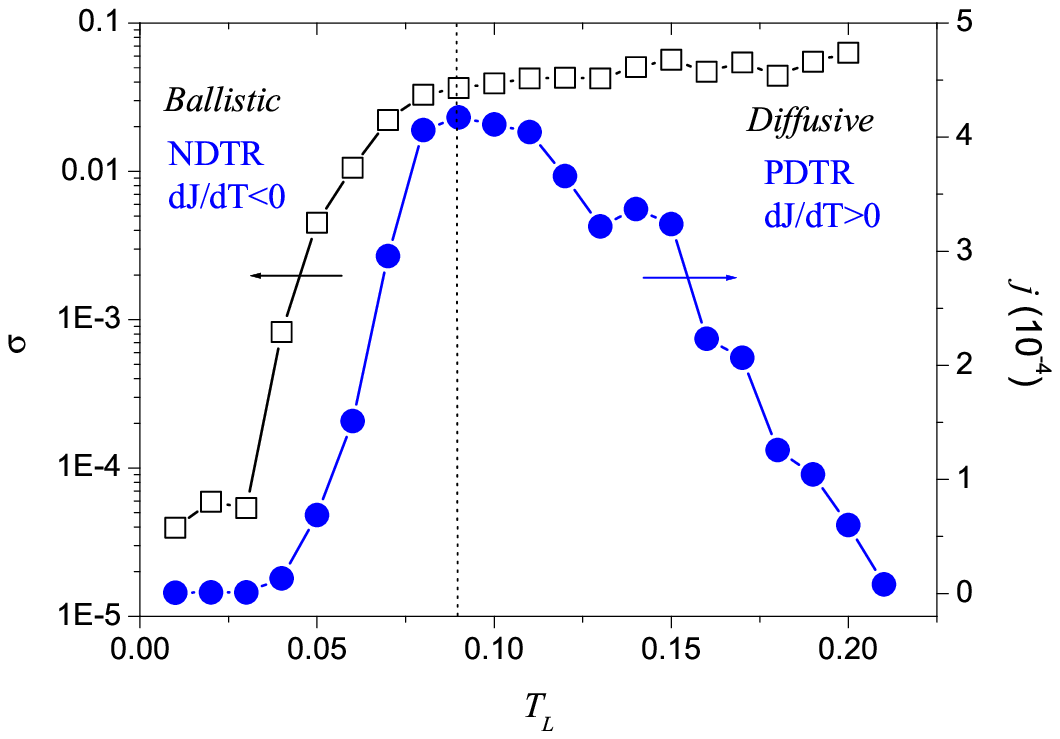';file-properties "XNPEU";}}

The phenomenon of NDTR can be understood from the theoretical approach. We
use Landauer equation to calculate the heat flux. For example, in the
ballistic regime the phonon heat flux can be also calculated through a
Landauer type expression \cite{Rego}: 
\begin{equation}
J=\int d\omega \omega \chi (\omega )[n_{R}(\omega )-n_{L}(\omega )],
\end{equation}%
$\allowbreak $where $\chi (\omega )$ is the temperature independent
transmission coefficient for phonons of frequency $\omega $. Here $%
n_{M}(\omega )=(e^{\beta _{M}\omega }-1)^{-1}$, $\beta _{M}=1/k_{B}T_{M}$, $%
(M=L,R)$ is the Bose-Einstein distribution characterizing the heat baths.

As reported in Ref.\cite{Segal2}, under the assumption of weak system-bath
interactions and when going into the Markovian limit, the probabilities $%
P_{n}$ to occupy the $n$ state of the phonon are found to satisfy the master
equation 
\begin{equation}
\overset{\cdot }{P_{n}}%
=(n+1)k_{d}P_{n+1}+nk_{u}P_{n-1}-[nk_{d}+(n+1)k_{u}]P_{n},
\end{equation}%
here the occupations are normalized $\sum P_{n}=1$, and $k_{d}$ and $k_{u}$
are the vibrational relaxation and excitation rates, respectively.

When going into weak system-bath interaction, $k_{d}$ and $k_{u}$ can
satisfy:%
\begin{equation}
k_{d}=k_{L}+k_{R},k_{u}=k_{L}e^{-\beta _{L}\omega _{0}}+k_{R}e^{-\beta
_{R}\omega _{0}},
\end{equation}%
with $k_{M}=\Gamma _{M}(\omega )[1+n_{M}(\omega _{0})]$, where $n_{M}(\omega
)=(e^{\beta _{M}\omega }-1)^{-1}$, $\Gamma _{M}(\omega )=\frac{\pi }{%
2m\omega ^{2}}\sum \alpha _{j}^{2}\delta (\omega -\omega _{j})$ and $\alpha
_{j}=\overset{-}{\alpha _{j}}\sqrt{2m\omega _{0}}$, here, $m$, $\omega _{0}$%
, $\overset{-}{\alpha _{j}}$are molecular oscillator mass, frequency and
coupling between the system and the heat baths, respectively.

The thermal properties of our model are obtained from the stationary state
solution of Eq.(4). The thermal flux is given by%
\begin{equation}
J=\omega _{0}\sum n(k_{L}P_{n}-k_{L}P_{n-1}e^{-\beta _{L}\omega _{0}}).
\end{equation}%
here positive sign denotes current flowing from right to left. In this
equation the first term indicates the thermal flux flowing from the L chains
into the L heat bath. The second term gives the oppositely flowing flux from
L heat bath to the chain. The thermal flux could be equivalently calculated
at the R chain.

So for the asymmetry system, which is similar to the case of a highly
anharmonic molecule coupled-possible asymmetrically-but linearly, to two
heat baths of different temperatures. Here, we simulate strong anharmonicity
by modeling the anharmonic two-segment chains by a two levels system that
notes a highly anharmonic vibrational mode \cite{Segal2}. The Hamiltonian
for this model is the same as presented in Eq.(1), except that we take $n=0,1
$ only. Following Eq.(3)- (6) and going into the classical limit, the heat
current reduces into the simple form%
\begin{eqnarray}
J &=&\omega _{0}\frac{\Gamma _{L}\Gamma _{R}(n_{R}-n_{L})}{\Gamma
_{L}(1+2n_{L})+\Gamma _{R}(1+2n_{R})} \\
&=&\frac{\Gamma _{L}\Gamma _{R}}{\Gamma _{L}+\Gamma _{R}}\frac{\omega _{0}}{%
T_{s}}(T_{R}-T_{L})[\exp (\omega _{0}/T_{s})+1]^{-1},
\end{eqnarray}%
in which%
\begin{equation}
T_{s}=\frac{\Gamma _{L}T_{L}+\Gamma _{R}T_{R}}{\Gamma _{L}+\Gamma _{R}}.
\end{equation}

We can now clearly seek the main factors which influence the heat current.
The thermal flux is given by multiplying four terms: (1) A symmetric
prefactor $\Gamma _{L}\Gamma _{R}/(\Gamma _{L}+\Gamma _{R})$. (2) The
characteristic frequency and effective temperature $\omega _{0}/T_{s}$. (3)
The temperature difference ($T_{R}-T_{L})$. (4) The molecular occupation
factor $[\exp (\omega _{0}/T_{s})+1]^{-1}$ \cite{Segal2}. This expression
denotes that the heat current is mainly a competitive effect between the
temperature difference and the molecular occupation factor. In the case of
low temperature (\emph{\emph{i}.e., }in the ballistic regime), the molecular
occupation factor changes obviously with $T_{s}$ by going into an
exponential function, while in the case of high temperature ($T_{s}$%
\TEXTsymbol{>}\TEXTsymbol{>}$\omega _{0}$, indicates the diffusive regime)
the molecular occupation factor is a constant $1/2$.

Figure 2 shows the simulated results of the temperature dependence of NDTR.
When $T_{R}$ is small, there exists NDTR. As $T_{R}$ increases, however, the
NDTR disappears. As also shown in Fig.3, a theoretical estimation Eq.(8)
based on Landauer equation confirms this temperature dependence of NDTR
simultaneously.\FRAME{ftbpFU}{4.8706in}{3.7766in}{0pt}{\Qcb{Heat current
ratio as a function of the temperature $T_{L}$ for $T_{R}$=0.11, 0.31, 0.51,
and 1.01. The system parameters are $V_{L}=5.0,$ $V_{R}=1.0,$ $K_{L}=1.0,$ $%
K_{R}=0.2,$ $Kint=0.05,$ and $N_{L}$=$N_{R}$=50. }}{}{fig2.eps}{\special%
{language "Scientific Word";type "GRAPHIC";maintain-aspect-ratio
TRUE;display "USEDEF";valid_file "F";width 4.8706in;height 3.7766in;depth
0pt;original-width 4.817in;original-height 3.7291in;cropleft "0";croptop
"1";cropright "1";cropbottom "0";filename '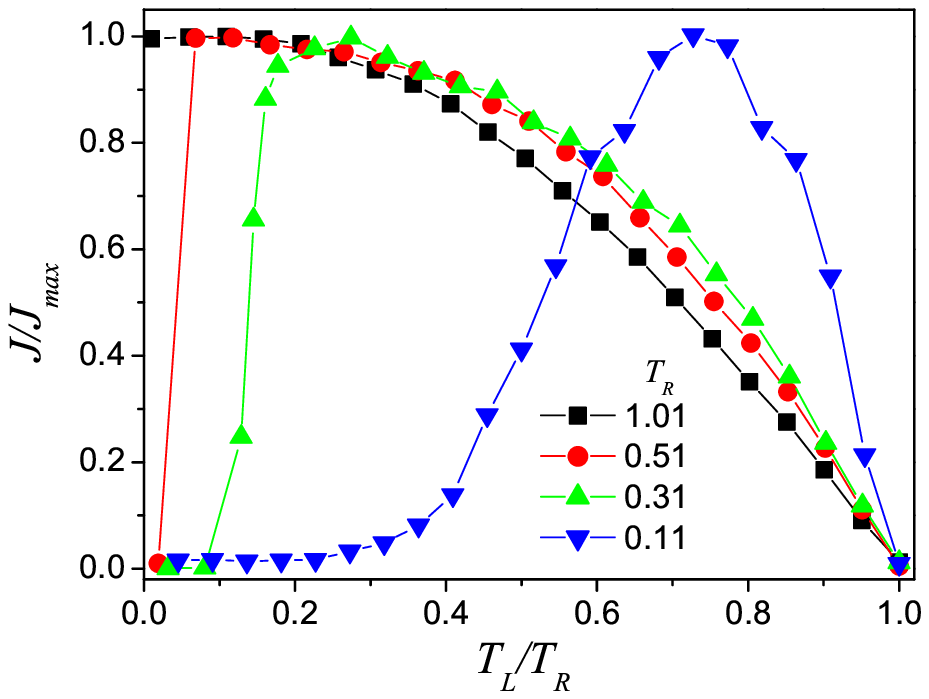';file-properties
"XNPEU";}}\FRAME{ftbpFU}{4.4149in}{3.2872in}{0pt}{\Qcb{The temperature
dependence of heat current. Results are obtained from Eq.(8) with $T_{R}$%
=200K (full), 100K (dash), 50K (dot). The remain parameters are $\protect%
\omega _{0}=150meV,$ $\Gamma _{L}=1.0meV,$ $\Gamma _{R}=1.4meV.$}}{}{fig3.eps%
}{\special{language "Scientific Word";type "GRAPHIC";display
"USEDEF";valid_file "F";width 4.4149in;height 3.2872in;depth
0pt;original-width 3.576in;original-height 3.787in;cropleft "0";croptop
"1";cropright "1";cropbottom "0";filename '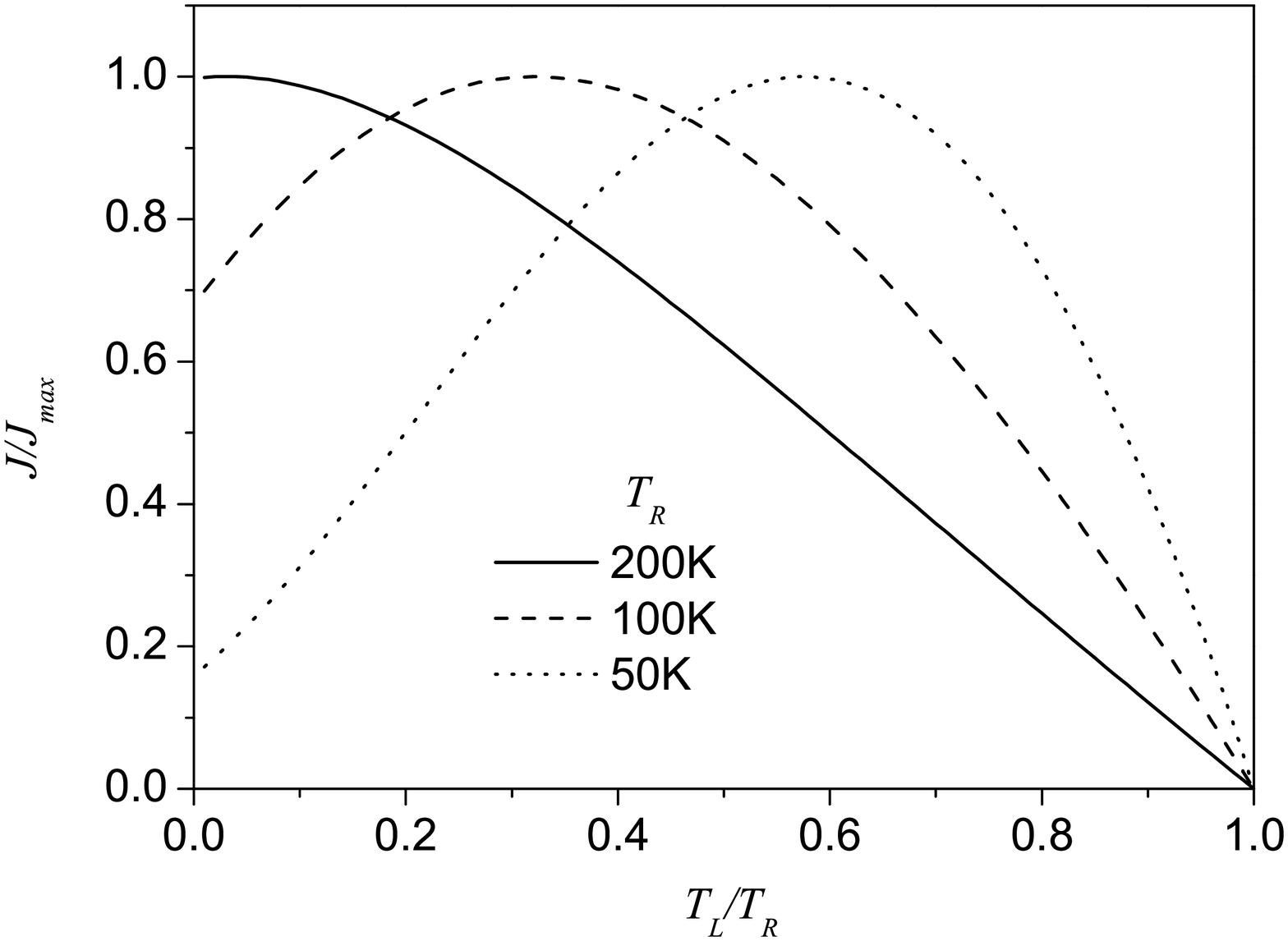';file-properties
"XNPEU";}}

Why\ does the NDTR appear at low temperature and disappear at high
temperature? This phenomenon can be also understood in detail from the
crossover from the diffusive to ballistic transport. In the case of low
temperature $T_{R}$, as shown in Fig.4a, when $T_{L}$ changes from $T_{R}$
to zero, the temperature profiles of the lattices illustrate a zero
temperature gradient. This diffusive-ballistic transition with temperature
difference is also shown in Fig.4b. When the temperature difference is
small, the thermal conductance does not change with $T_{L}$. However, when
the temperature difference is large, the thermal conductance decreases
linearly with $T_{L}$ decreasing. It can be interpreted that the
diffusive-ballistic transition induces the NDTR, which takes place just when
the ballistic transport prevails over the diffusive transport. In the case
of high temperature $T_{R}$, even when $T_{L}$ is very small, the
temperature gradient are non-zero and the thermal conductance does not
change with $T_{L}$. There exists no diffusive-ballistic transition.
Therefore, the NDTR does not occur in the case of high temperature $T_{R}$. 
\FRAME{ftbpFU}{3.6564in}{3.7844in}{0pt}{\Qcb{(a) Temperature profiles of the
lattices for different heat baths for $T_{L}$=0.02. (b) Thermal conductance
as a function of the temperature $T_{L}/T_{R}$ for different temperatures $%
T_{R}$=0.11, 0.21, 0.41, 0.81 and 1.01 (from bottom to top). Remain
parameters are the same as for Fig.2.}}{}{fig4.eps}{\special{language
"Scientific Word";type "GRAPHIC";maintain-aspect-ratio TRUE;display
"USEDEF";valid_file "F";width 3.6564in;height 3.7844in;depth
0pt;original-width 3.6097in;original-height 3.7369in;cropleft "0";croptop
"1";cropright "1";cropbottom "0";filename '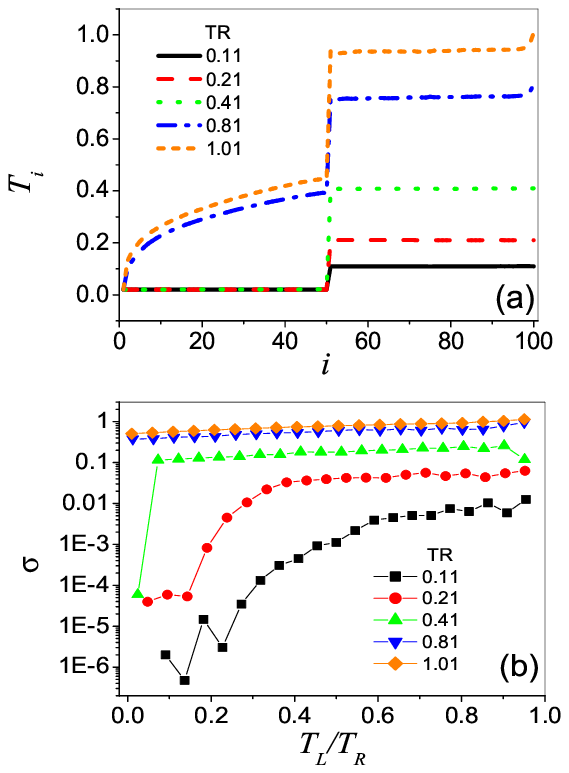';file-properties
"XNPEU";}}

We would also like to discuss the finite size effect of NDTR. As shown in
Ref. \cite{ZGShao}. When the system size increases, the phenomenon of NDTR
disappears. It is interesting to present the system size dependence of
thermal conductance, which is shown in Fig.5. In the ballistic regime (\emph{%
\emph{i}.e.,} when the system size is much smaller than the phonon mean free
path), thermal conductance increases linearly with the system size. In the
diffusive regime (\emph{\emph{i}.e.,} when the system size increases to far
larger than the phonon mean free path), thermal conductance will be
independence of the system size. There exists a crossover from ballistic to
diffusive transport with the increase of the system size. When the system
goes to completely diffusive transport regime, the temperature difference
does not change the transport mode and then the NDTR disappears.\FRAME{ftbpFU%
}{4.395in}{3.3157in}{0pt}{\Qcb{Thermal conductance versus lattice length $%
N(=N_{L}+N_{R})$ for $T_{R}=0.21$ and $T_{L}=0.05$ (Triangles), $0.07$%
(Circles), and $0.15$(Squares). Remain parameters are the same as for Fig.2.}%
}{}{fig5.eps}{\special{language "Scientific Word";type
"GRAPHIC";maintain-aspect-ratio TRUE;display "USEDEF";valid_file "F";width
4.395in;height 3.3157in;depth 0pt;original-width 4.3439in;original-height
3.2707in;cropleft "0";croptop "1";cropright "1";cropbottom "0";filename
'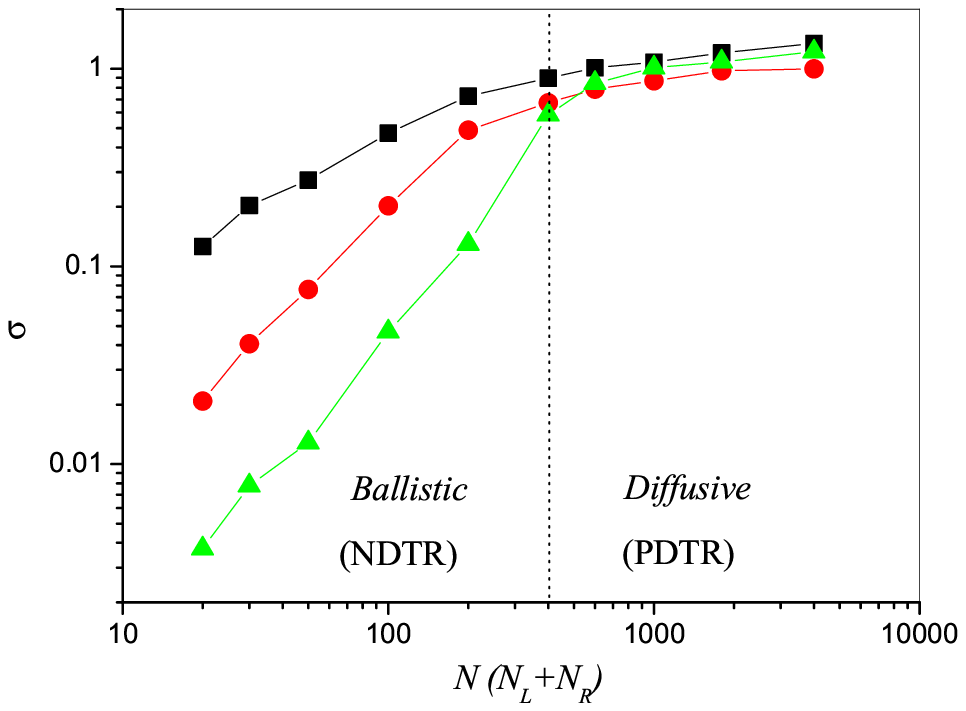';file-properties "XNPEU";}}

In conclusions, we have found the physical mechanism of the negative
differential thermal resistance in two-segment asymmetric F-K chains through
both theoretical analyses and numerical simulations. From the analytical
relationship between the heat current and the temperature, the NDTR effect
can be understood from a competition between the temperature difference and
the molecular occupation factor. In the ballistic regime the molecular
occupation factor mainly depends seriously on the temperature. However, in
the diffusive regime, the molecular occupation factor is a constant. The
NDTR effect occurs just because the decrease of the molecular occupation
factor is faster than the increase of the temperature difference. The main
factors that influence the transport modes of the phonons are the
temperature and the system size. We propose that different system structures
can affect the temperature dependence of NDTR through changing the
transmission mode of the phonons. When there exists a crossover from
diffusive to ballistic transport, the NDTR can be observed, otherwise it
will not. Due to the rigorous condition of validity: low temperature and
small system, we still have a long way to go before the NDTR effect is
produced in fabricated materials.

\begin{acknowledgments}
We would like to thank members of the Centre for Nonlinear Studies for
useful discussions. This work was supported in part by grants from the Hong
Kong Research Grants Council (RGC) and the Hong Kong Baptist University
Faculty Research Grant (FRG).
\end{acknowledgments}


\begin{thebibliography}{99}
\bibitem{B.Li1} L. Wang, and B. W. Li, Phys. Rev. Lett. \textbf{99}, 177208
(2007); B. Li, Lei Wang, and Giulio Casati, Appl. Phys. Lett. \textbf{88},
143501 (2006).

\bibitem{B.Li3} W. C. Lo, L. Wang, and B. Li, J. Phys. Soc. JPN \textbf{77},
054402 (2008).

\bibitem{B.Li4} B. Li, L. Wang, and G. Casati, Phys. Rev. Lett. \textbf{93},
184301 (2004).

\bibitem{Segal2} D. Segal, Phys. Rev. B \textbf{73}, 205415 (2006).

\bibitem{Bardeen} J. Bardeen and W. H. Brattain, Phys. Rev. \textbf{74}, 230
(1948).

\bibitem{Segal1} D. Segal, Phys. Rev. E \textbf{77}, 021103 (2008).

\bibitem{CWChang} C. W. Chang, D. Okawa, A.Majumdar, and A. Zettl, Science 
\textbf{314}, 1121 (2006).

\bibitem{YZhang} B. Hu, L. Yang, and Y. Zhang, Phys. Rev. Lett. \textbf{97},
124302 (2006).

\bibitem{ZGShao} Z. G. Shao, L. Yang, and B. Hu, Transition From the
Negative Differential Thermal Resistance to the Ordinary State, ReportNo.
CNS-08-20, 2008.

\bibitem{Ai} B. Q. Ai, Negative differential thermal resistance in
sysmmetric F-K single-chains, ReportNo. CNS-10-3, 2008.

\bibitem{Phonon} J. A. Reissland, The Physics of Phonons (John Wiley \& Sons
Ltd., London, 1973).

\bibitem{SolidPhysics} J. R. Christman, Fundamentals of Solid State Physics
(John Wiley \& Sons Ltd., New York, 1988).

\bibitem{Roy} D. Roy, Phys. Rev. E \textbf{77}, 062102 (2008).

\bibitem{Nose} S. Nose, J. Chem. Phys. \textbf{81}, 511 (1984); W. G.
Hoover, Phys. Rev.A \textbf{31}, 1695 (1985).

\bibitem{Art} D. C. Rapaport, The Art of Molecular Dynamics Simulations,
(Cambridge University Press, Cambridge, 2001).

\bibitem{Press} W. H. Press, S. A. Teukolsky, W. T. Vetterling, and B. P.
Flannery, Numerical Recipes (Cambridge University Press, Cambridge, 1992).

\bibitem{JSWang1} J. S. Wang, Phys. Rev. Lett. \textbf{99}, 160601 (2007).

\bibitem{JSWang2} Y. Xu, J. S. Wang, W. Duan, B. L. Gu, and B. Li,
arXiv:0807.3819v1 (2008).

\bibitem{DeyuLi} D. Li, Y. Wu, P. Kim, L. Shi, P. Yang, and A. Majumdara,
Appl. Phys. Lett. \textbf{83}, 2934 (2003).

\bibitem{Livi} S. Lepri, R. Livi, and A. Politi, Phys. Rep. \textbf{377}, 1
(2003).

\bibitem{Rego} L. G. C. Rego, G. Kirczenow, Phys. Rev. Lett \textbf{81}, 232
(1998).
\end{thebibliography}
\end{document}